# Quantum Computing in non Euclidean Geometry

Germano Resconi
Catholic University, via Trieste 17 Brescia, Italy
resconi@numerica.it

Ignazio Licata
ISEM, Inst. For Scientific Methodology, PA, Italy
Ignazio.licata@ejtp.info

## Abstract

The recent debate on hyper-computation has raised new questions both on the computational abilities of quantum systems and the Church-Turing Thesis role in Physics We propose here the idea of "geometry of effective physical process" as the essentially physical notion of computation. A key element to understand the shape of the geometry in Quantum mechanics is the Fisher metric which meaning is connected with the stabilisation of the Classical mechanics by Quantum mechanics. Different researchers define the stabilisation as the manifestation of the dissipative process into the vacuum. We argue that no dissipation takes place, but a more general change occurs in Quantum mechanics. In Quantum mechanics we cannot use the traditional Euclidean geometry but we introduce more sophisticate non Euclidean geometry which include a new kind of information diffuse in  the entire universe and that we can represent as Fisher information or active information. We remark that from the Fisher information we can obtain the Bohm and Hiley quantum potential and the classical Schrödinger equation. The bridge between Quantum mechanics and Classical mechanics is given by the Fisher metric in statistic geometry which value is obtained by Quantum mechanics. We can see the quantum phenomena do not affect a limited region of the space but is reflected in a change of the geometry of all the universe. In conclusion any local physical change or physical process is reflected in all the universe by the change of its geometry, This is the deepest meaning of the entanglement in Quantum mechanics and quantum computing. We stress the connection between metric and information as measure of change. We analyze how the standard form (quantum gates) and the non-standard form of the quantum computing can be seen as a particular case of the metric of the parameters space in the distribution of the probabilities. Because computation is not restricted to calculus but is  the environment changing via physical processes, super-Turing potentialities derive from an incomputable information source embedded into the geometry of the universe in accordance with Bell's constraints. On condition that we consider the formal concept of "universality" as a particular case of the universal geometry of the probabilistic space with its transformations we open the possibility that quantum oracles can be reachable. In this way computation is led back to the hidden universal geometry of  the physical world. In the general relativity we define the geometry of the space time. In our approach quantum phenomena define the geometry of the parameters of the probability distribution that include also the space time parameters. To study this new approach to the computation we use the new theory of Morphogenic systems.

## 1. Computation, Physics and Geometry: an Unitary Perspective

Processing information is what all physical systems do. Such intuition first expressed by Rolf Landauer (Landauer, 1991; 1996) with exemplary clarity has recently risen from the ranks giving birth to extremely interesting and promising developments.

The latest computational models have increasingly undermined the privileged position of Turing-Computation model and the role of Church-Turing thesis, as well. The various kinds of Unconventional Computing focus on either different vocations than Turing-Machine, such as the attention for the spatial and temporal features of computing, or schemes of information processing related to refined forms of non-linearity, fuzziness and infinite or non-computable many values (C. Teuscher et al. 2008).

These new models are connected one with the others by a more or less radical breaking between the richness of the information processes supported by a great variety of physical systems and the classical computation model which appears so to be limiting in some aspects and too indefinite in some others. After all, it is just for such features that Turing Computation has been so successful and its characteristics are regarded as "universal". This gap pushed the researchers to propose computational models where the effective calculability of a function is redefined in terms of evolutionary dynamics of the physical system's configurations, and thus the computational machinery is the system itself under particular constraints (for the definition of "effective physical process" see Licata, 2007).

Thus, these are contextual and collective kinds of computation, where the resources of a system (values of the observables, quantum states, "molecules", agents, etc.) give birth to cooperative processes from which the realization of a task not referable to a recursive function emerges. Besides, it is easy to demonstrate that simple systems of interacting Turing-Machine can show hyper-computational abilities (Kieu and Ord, 2005). In other words, computation is strongly tied to the very physical nature of the system and its global configuration, and the "algorithm" is the evolution itself of the system.

The problem of an alternative model becomes crucial in quantum computing. It is known that a Quantum Turing Machine can be formally defined so as to extend the classical paradigm to the calculation with qbit (see, for example, Perdrix and Jorrand, 2006). The outcome is however controversial: within such scheme Quantum Computation does not seem more powerful, but only more effective. And more: in some cases it is possible to demonstrate that the performances of the Quantum computing Turing scheme-based can be obtained also by classical systems in polynomial time (Ahronov, 2007; Calude, 2007).

The recent works on Adiabatic Quantum Computation and Quantum Neural Networks thus suggest that a model for "Schrödinger Machine" (Milbur, 1997) has to be searched in a different direction as well as the classical paradigm appears as a cage for the computational potentialities of Quantum Physics.

Patently, Church-Turing thesis cannot be extended to such kinds of computation, and it is so necessary to point our attention to a different framework which can provide us with a unitary vision of the "unconventional" kinds of computation. The Theory of Morphogenic systems allows to naturally connecting the information a system can process with its geometric structure (Resconi and Nikravesh, 2007).

The basic idea can be directly traced back to General Relativity philosophy, where the principle of general covariance is substantially a principle of conservation of information and the dynamics of the gravitational field is locally individuated by its geometry. As for morphogenic systems - within which the different known forms of classic and quantum computation are included - the metrics of the system's geometry characterizes the peculiar way how information is processed. It is physically significant noticing that the geometry emerging from morphic computing is non-Euclidean, because it directly derives from a generalization of the notion of "field". Such thing mirrors the key idea of our work, and it is also the conceptual connecting point for the different forms of computation, which is to say that the informational activity of a system is strongly connected to its morphogenic field configurations. In this way, the tensor calculus appears as an extremely useful tool to define the different computational models and the classical notion of "universality" is replaced by the specific metrics of each system. We can also state that the non-Euclidean aspects describe the way how the computational model differs from the classical Turing one, which is based on local algebraic operations and indifferent to the notion of field.

The quantum case has to be dealt with in different way. Here an "active information" field (Bohm and Hiley, 2005) makes its appearance; it has no equivalent in Classical Physics and indicates the non-local features of quantum domain. The most suitable morphogenic system for quantum computing is built on the parametric space of probability distribution. We will show that the Fisher information role can directly be linked to the non-locality and the anticommutativity of quantum information and is not connected to any specific interpretation.

The notion of geometry has also a significance directly connected to the task: computation – considered as an activity oriented by an observer – depends on the adopted experimental configuration, and the hypercomputational potentialities – far from being limit situations only occurring in exotic physical environments – depend upon the transformations of the system's geometry. Consequently the term "programming" takes on a completely new meaning just in relation to the particular geometry the experimental apparatus defines. In the same way as Gödel outcomes are considered "limiting" within the Hilbert axiomatic program, whereas they reveal the open logic of mathematics if regarded from a more general viewpoint (Chaitin, 2006), the super-Turing possibilities of oracles emerge from a vision which links geometry to information.

The paper arrangement is: in paragraph 2 the concept of morphogenic system is introduced; in paragraph 3 the geometric image of projection operator is studied; paragraph 4 is dedicated to the essential features of morphic computing and tensor calculus (invariants and Lagrangian Function Minimum Condition) is introduced. Finally, paragraph 5 is dedicated to Quantum Morphic Computation and to the relations between Fisher information and anticommutativity.

## 2. Morphogenic Systems

### 2.1. Conventional and Unconventional Computations

Software is built for a specific purpose, with the actions (or operations) of software containing a set of instructions to be implemented in the context of the required purpose or task. Traditionally designed and implemented software uses a standard programming language that contains the syntactic rules that aggregate a set of instructions. The Turing machine is used as a conceptual model and system architectures are based on the same concepts to realise the purpose. However, observations in nature show that the purpose is obtained without taking into consideration of the Turing Machine (TM) model. In the proposed approach, we suggest that biomimetics can be used for the construction of software prototypes as a general computation model. At any time in the TM model, we follow a rule to decide how to take actions in a given moment and in one specific location of the memory. Therefore, in one table we have to decide on all possible actions, as well as the path (trajectory) of actions resulting in the implementation of a required task. Considering that no external assistance exists in defining a purpose, nor a predefined plan for a desired path of actions to generate the purpose, only conceptual work can provide a definition of the path. Every element of the path is independent from any other elements, and the connection is made at the conceptual level with user intervention. The Turing Machine only offers the connection between any element of the path and the actions, so complex actions have to be defined by a path of elementary actions. In fact, the semantic part of the purpose is forgotten and only remains in the conceptual mind of the developer. This situation creates hardship to the developer, as they cannot take over the control of the computation process without an extensive analysis of the problem and build a suitable execution path to obtain required outcome. Only the arrival of a new generation of programming language that does not adhere to the primitive TM model can allow a definition of flexible software prototypes, in which a path is seen as a variational or extremum principle, which as a consequence has significant improvements to traditional software practices. It is clear that natural languages are beyond the Turing Machine. We think that TM is one of the models used as a starting point for any conceptual improvement to computational methodologies.

Recent work in the domains of autopoietic systems, biomimetic middleware systems, constructal theory, immuno-computing, holographic computing, morphic computing, quantum computers, DNA computing, secondary sound sources, General system logical theory, Neural Networks and so forth suggest a change is required to the traditional approaches based on the original Turing Machine model ( Calude et al., 1998). The first point is the new computation model offers a total change of perspective in development. By beginning to state the purpose as the starting point for defining the computation process, the purpose becomes the conceptual input to a computer or a machine. Secondly, one should ignore the local machine state and generate the context and its rules as resources to locate or allocate the computational component(s).

The introduction of the concept of a context dramatically changes the definition of self in the Turing Machine. The TM states that various actions and processes that use memory are perceived as separate entities and the self is disconnected from individual entities. In principle, the actions in the TM are coordinates defined by a table of state changes. The table is defined by an external entity, typically a human, and thus is justified by a given requirement to execute a specific purpose. In the Turing Machine, the table of the state changes is defined as an axiom without any explanation or computation model. In similar manner, software is built as an axiom without any explanation or computational reference. As a result, in the TM we cannot execute software for the famous Undecidable Theories. When proposing the new computational model, we extend the definition of self to a set of entities that are strongly inter-connected or correlated. Hence, the first definition of the global self entity is the context. The context is not created by just connecting individual entities with the individual self. Self is associated with all elements of the context at once. The same stands for the definition of Purpose which in the global (morphogenic) computation is defined by a task or a goal made by a set of conceptual entities that cannot be separated and that are associated with one self only, thus, individual entities become one with the proper self. An example of such a morphogenic computation can be found in the Dempster-Shafer Evidence theory where a set of inseparable elements can be measured (G. Resconi, G. J. Klir; E. Pessa, 1999). A similar example can be found in quantum mechanics where the entanglement phenomena change the definition of a particle perceived as a self into a set of particles entangled or correlated with only one self or with only one particle. In quantum mechanics, when the state of one particle is changed, it is correlated with others. Thus, in one instance of change in one particle, the state of all other correlated particles changes as well. In the quantum computer, the superposition quantum phenomena can be observed, where in one location or time, millions of states can be superimposed forming one super or meta-state. In the meta-state we only have one self in which all states cannot be separated one from another as they are all part of the same meta-state. In the immuno-computing, the conceptual binding phenomena link the biological entities in one self (Tarakanov et al., 2003). Similarly, the Hopfield network of neurons joins as one form of computational energy in a multi-dimensional function or self (Ripley, 2008).

Similarly to holography, for the morphogenic system the global operations of *write* and *read* play an important role. The Constructal Theory and allometric rules that define the concepts of self and self-shaping in particular are vital to understand the key concepts (Bejan & Lorente, 2008). The constraint is a purpose in dissipative thermodynamics, where the variational and extremum principle realises the context of purpose or constraint. In the quantum global phenomena, the computation is associated with two main operations: Unitary Transformation (UT) and the Projection Operator. The Unitary Transformation changes all states of the system at the same time and as the self. The main motivation for defining Unitary Transformation is to be able connect invariants of physical quantum dynamics with the associated superposition of states while the chief reason for defining the Projection Operator is to be able to associate space vectors with the quantum measure thus allowing collapse of the system to an individual state. Additionally, Projection Operator allows explaining these complex operations and processes in very simple terms. In fact, we can distinguish physics from other scientific professions, as it is the study of being able to reduce a given system to its most significant subsystem components, instead of taking into consideration all details of a system.

## 2.2 Computation in Morphogenic Systems

In the morphogenic system, we have an abstract universe of the objects. Now the universe of the objects has two interpretations as in the case of the voltages and currents in the electrical circuit. For the space of the voltages the objects are the voltages at any edge of the electrical circuit. For the space of the currents the objects are the currents in any edge. The dimension of the object space is equal to the number of edges in the electrical circuit. Other possible dual variables can be used in the morphogenic system as forces and the fluxes in mechanics or dissipative thermodynamics; in general the dual interpretation of the object space will be denoted as causes and effects. The morphogenic system is modeled by samples of the causes and effects. Samples are vectors or points in the space of the objects. Given the samples of the effect and the purpose as a virtual cause, we compute the vector E of the scalar products of the purpose with the vectors of the samples of the effects. The vector E is the effective origin of the causes. For example, when the purpose is the virtual voltages in the electrical circuit, E are the voltages sources in the mesh of the electrical circuit. With the relation between cause and effects we compute the independent effects I of the computed causes E. For electrical circuit I are the currents in any mesh of the circuit. The values I are the intensity of the effect samples which superposition C gives the effects coherent with given purpose. With the effect/cause rule we can compute the effective causes that give the effective effect coherent with the samples.

In conclusion from the virtual cause given by purpose we generate the effective causes in agreement with the samples. The algorithm that we describe is denoted as the projection operator that transforms a virtual cause (purpose) into an effective cause. In the electrical circuit given a virtual set of voltages (purpose) we compute E as the generators of the voltage, I as the currents in the mesh, C as the currents in any part of the circuit and at the end the voltages in the electrical circuit as the effective cause of the currents by the Ohm's law. Given I and E vectors, the scalar product P of the two vectors takes the minimum value for computed values of I and E obtained by the purpose vector. The projection operator gives the solution of the minimum value of the scalar product of I and E with the given vector E. For the electrical circuit the relation between the voltages (cause) and the currents (effects) is Ohm's rule by impedances. The scalar product of E and I is the power of the electrical circuit. The current fluxes in the circuit under the action of the source of the voltages E so as to obtain the minimum dissipation of the power. The space of the samples of causes and effects are non Euclidean spaces which metric is P. The cross product of the samples cause effect defines the geometry of the sample space. The unitary transformation U of cause E and effect I gives a new form of the metric with the same minimum value of P. In the electrical circuit we can change the currents and the impedances so as to have always the same power. The property of the unitary matrix is comparable with the minimum action and symmetry in mechanics and also the unitary transformation in the quantum mechanics with the same probability.

With electrical analogy we remark that given in input wanted vector of voltages or stimulus, we can design suitable currents or response to obtain the best voltages approximation to the wanted task or voltages inside the physical electrical circuit. Because from voltages we came back to the voltages again the operator that we use is a loop in the network of morphogenic system where we locate the suitable operators or context.

Given the cause as input and effect as output

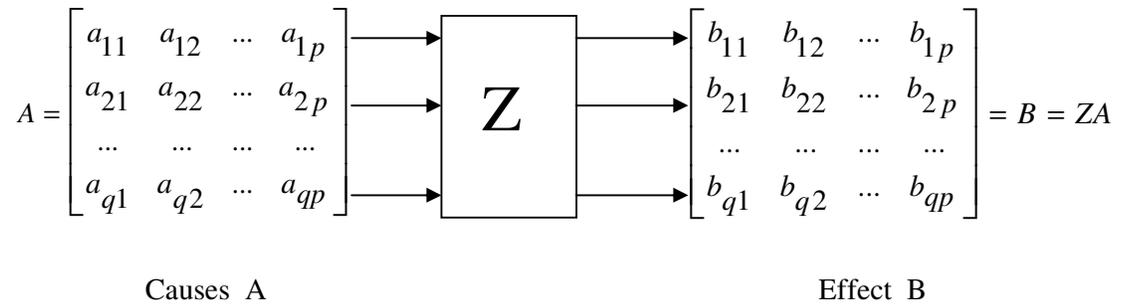

Causes  A                                                    Effect  B

Figure 1 Cause effect system (MIMO many inputs/ many outputs system). In input we have p strings with q values. The same in output. The matrix Z makes the transformation.

The matrix A is the matrices of the p samples for q inputs of the causes

$$A = \begin{bmatrix} a_{11} & a_{12} & \cdots & a_{1p} \\ a_{21} & a_{22} & \cdots & a_{2p} \\ \cdots & \cdots & \cdots & \cdots \\ a_{q1} & a_{q2} & \cdots & a_{qp} \end{bmatrix}$$

For the samples of output or effect we have

$$B = ZA = \begin{bmatrix} b_{11} & b_{12} & \cdots & b_{1p} \\ b_{21} & b_{22} & \cdots & b_{2p} \\ \cdots & \cdots & \cdots & \cdots \\ b_{q1} & b_{q2} & \cdots & b_{qp} \end{bmatrix}$$

Now B and A can be connected by the matrix:

$$Z = B \ (A^T \ A)^{-1} \ A^T$$

In fact we have

$$Z \ A = B \ (A^T \ A)^{-1} \ A^T \ A = B$$

Now given A and B we can also generate a simple loop in this way

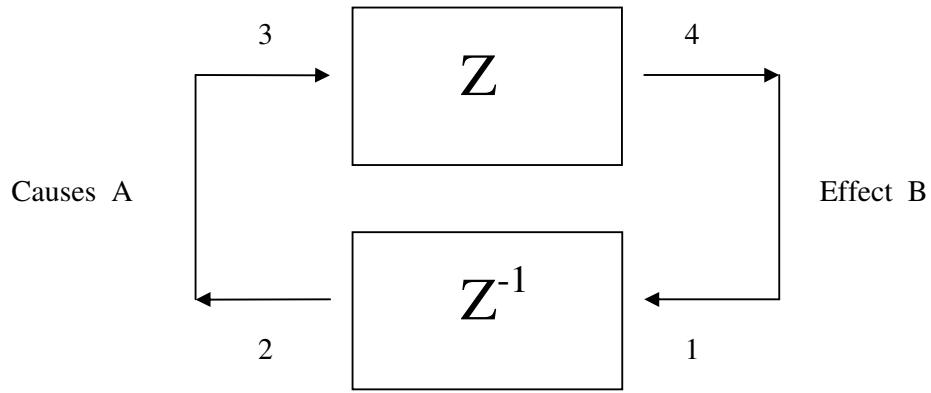

Figure 2  Morphogenic loop

For the definition of Z in the previous case we have

$$Z = B \ (A^T \ A)^{-1} \ A^T$$

$$Z^{-1} = A \ (B^T \ B)^{-1} \ B^T$$

and

$$Z \ A \ = B \ \text{and} \ Z^{-1} \ B = A$$

So for the loop in figure 2 we have

$$Q = Z\,Z^{-1} = B\,(A^T A)^{-1}\,A^T A\,(B^T B)^{-1}\,B^T = B\,(B^T B)^{-1}\,B^T$$

Where $Q\,B = B$ , and

$$Q^2 = B\,(B^T B)^{-1}\,B^T B\,(B^T B)^{-1}\,B^T = B\,(B^T B)^{-1}\,B^T = Q$$

So Q is a projection operator. For the projection operator we have another form. In fact we can write

$$Q = Z^{-1}\,Z = A\,(B^T B)^{-1}\,B^T B\,(A^T A)^{-1}\,A^T = A\,(A^T A)^{-1}\,A^T$$

Where $Q\,A = A$ , and

$$Q^2 = A\,(A^T A)^{-1}\,A^T A\,(A^T A)^{-1}\,A^T = A\,(A^T A)^{-1}\,A^T = Q$$

Now for Z we have another representation

$$Z = B\,(B^T A)^{-1}\,B^T$$

$$Z^{-1} = A\,(A^T B)^{-1}\,A^T$$

And

$$Q = Z^{-1}\,Z = A\,(A^T B)^{-1}\,A^T B\,(B^T A)^{-1}\,B^T = A\,(B^T A)^{-1}\,B^T$$

Where

$$Q\,A = A \quad , \quad Q^2 = A\,(B^T A)^{-1}\,B^T A\,(B^T A)^{-1}\,B^T = A\,(B^T A)^{-1}\,B^T = Q$$

We have also

$$Q = Z\,Z^{-1} = B\,(B^T A)^{-1}\,B^T A\,(A^T B)^{-1}\,A^T = A\,(A^T B)^{-1}\,A^T$$

Where

$$Q\,B = B \quad , \quad Q^2 = A\,(B^T A)^{-1}\,B^T A\,(B^T A)^{-1}\,B^T = A\,(B^T A)^{-1}\,B^T = Q$$

Now because

$$B = \begin{bmatrix} Z_{1,1} & Z_{1,2} & ... & Z_{1,q} \\ Z_{2,1} & Z_{2,2} & ... & Z_{2,q} \\ ... & ... & ... & ... \\ Z_{q,1} & Z_{q,2} & ... & Z_{q,q} \end{bmatrix} \begin{bmatrix} a_{11} & a_{12} & ... & a_{1p} \\ a_{21} & a_{22} & ... & a_{2p} \\ ... & ... & ... & ... \\ a_{q1} & a_{q2} & ... & a_{qp} \end{bmatrix} = \begin{bmatrix} \sum_j Z_{1,j} a_{j1} & \sum_j Z_{1,j} a_{j2} & ... & \sum_j Z_{1,j} a_{jp} \\ \sum_j Z_{2,j} a_{j1} & \sum_j Z_{2,j} a_{j1} & ... & \sum_j Z_{2,j} a_{jp} \\ ... & ... & ... \\ \sum_j Z_{q,j} a_{j1} & \sum_j Z_{q,j} a_{j1} & ... & \sum_j Z_{q,j} a_{jp} \end{bmatrix}$$

$$\begin{bmatrix} b_{11} & b_{12} & ... & b_{1p} \\ b_{21} & b_{22} & ... & b_{2p} \\ ... & ... & ... & ... \\ b_{q1} & b_{q2} & ... & b_{q,p} \end{bmatrix}$$

we have the q . p equations as constrain for the matrix Z. The free values for Z are q . q − q .p .

For example when p = 2 and q = 3 , we have 2 .3 = 6 values of Z fixed and 3 . 3 − 2 .3 = 3 free value for the Z. Now because we have

$$Z = B \ (B^T \ A)^{-1} \ B^T \ \text{ or } \ Z = B \ (A^T \ A)^{-1} \ A^T$$

We can fix the 6 values with the computation of Z and we are free to put the other values as we want. For example given

$$A = \begin{bmatrix} 1 & 0 \\ 1 & 1 \\ 0 & 1 \end{bmatrix} \text{ and } B = \begin{bmatrix} 1 & 0 \\ 1 & 1 \\ 1 & 0 \end{bmatrix}$$

We have

$$\begin{bmatrix} Z_{1,1}a_{1,1}+Z_{1,2}a_{2,1}+Z_{1,3}a_{3,1} & Z_{1,1}a_{1,2}+Z_{1,2}a_{2,2}+Z_{1,3}a_{3,2} \\ Z_{2,1}a_{1,1}+Z_{2,2}a_{2,1}+Z_{2,3}a_{3,1} & Z_{2,1}a_{1,2}+Z_{2,2}a_{2,2}+Z_{2,3}a_{3,2} \\ Z_{3,1}a_{1,1}+Z_{3,2}a_{2,1}+Z_{3,3}a_{3,1} & Z_{3,1}a_{1,2}+Z_{3,2}a_{2,2}+Z_{3,3}a_{3,2} \end{bmatrix} = \begin{bmatrix} b_{1,1} & b_{1,2} \\ b_{2,1} & b_{2,2} \\ b_{3,1} & b_{3,2} \end{bmatrix}$$

in our case we have

$$\begin{bmatrix} Z_{1,1}+Z_{1,2} & Z_{1,2}+Z_{1,3} \\ Z_{2,1}+Z_{2,2} & Z_{2,2}+Z_{2,3} \\ Z_{3,1}+Z_{3,2} & Z_{3,2}+Z_{3,3} \end{bmatrix} = \begin{bmatrix} 0 & 1 \\ 1 & 1 \\ 1 & 0 \end{bmatrix}$$

And the equations

$$\begin{bmatrix} Z_{1,1}+Z_{1,2}=0 & Z_{1,2}+Z_{1,3}=1 \\ Z_{2,1}+Z_{2,2}=1 & Z_{2,2}+Z_{2,3}=1 \\ Z_{3,1}+Z_{3,2}=1 & Z_{3,2}+Z_{3,3}=0 \end{bmatrix}$$

Now we can compute the impedance by the expression $Z = B\ (A^T\ A)^{-1}\ A^T$

$$Z = B(A^T A)^{-1} A = \begin{bmatrix} -\dfrac{1}{3} & \dfrac{1}{3} & \dfrac{2}{3} \\[2mm] \dfrac{1}{3} & \dfrac{2}{3} & -\dfrac{1}{3} \\[2mm] \dfrac{2}{3} & \dfrac{1}{3} & -\dfrac{1}{3} \end{bmatrix}$$

Now with the solution of Z, where all the previous 6 equations are true, we can generate all possible solutions in this way

$$\begin{bmatrix} Z_{1,1}+\alpha+Z_{1,2}-\alpha=0 & Z_{1,2}-\alpha+Z_{1,3}+\alpha=1 \\ Z_{2,1}+\beta+Z_{2,2}-\beta=1 & Z_{2,2}-\beta+Z_{2,3}+\beta=1 \\ Z_{3,1}+\gamma+Z_{3,2}-\gamma=1 & Z_{3,2}-\gamma+Z_{3,3}+\gamma=0 \end{bmatrix}$$

And from the previous values of Z we have:

$$Z = \begin{bmatrix} -\dfrac{1}{3}+\alpha & \dfrac{1}{3}-\alpha & \dfrac{2}{3}+\alpha \\[2mm] \dfrac{1}{3}+\beta & \dfrac{2}{3}-\beta & -\dfrac{1}{3}+\beta \\[2mm] \dfrac{2}{3}+\gamma & \dfrac{1}{3}-\gamma & -\dfrac{1}{3}+\gamma \end{bmatrix}$$

Now we choose the parameters so as to maximise the zero value in the matrix. For

$$\alpha = \frac{1}{3}, \beta = -\frac{1}{3}, \gamma = \frac{1}{3}$$

we have

$$Z = \begin{bmatrix} 0 & 0 & 1 \\ 0 & 1 & 0 \\ 1 & 0 & 0 \end{bmatrix}$$

That is the minimum representation for Z.

## 3. Geometric image of the projection operator

Given the matrix

$$H = \begin{bmatrix} h_{11} & h_{12} & ... & h_{1p} \\ h_{21} & h_{22} & ... & h_{2p} \\ ... & ... & ... & ... \\ h_{q1} & h_{q2} & ... & h_{qp} \end{bmatrix}$$

We have

$$\boldsymbol{Y} = HW = \begin{bmatrix} h_{11} & h_{12} & ... & h_{1p} \\ h_{21} & h_{22} & ... & h_{2p} \\ ... & ... & ... & ... \\ h_{q1} & h_{q2} & ... & h_{qp} \end{bmatrix} \begin{bmatrix} w_1 \\ w_2 \\ ... \\ w_p \end{bmatrix} = \begin{bmatrix} y_1 \\ y_2 \\ ... \\ y_q \end{bmatrix}$$

*and*

$$\boldsymbol{Y} = w_1 \begin{bmatrix} h_{11} \\ h_{21} \\ ... \\ h_{q1} \end{bmatrix} + w_2 \begin{bmatrix} h_{12} \\ h_{22} \\ ... \\ h_{q2} \end{bmatrix} + ..... + w_p \begin{bmatrix} h_{1p} \\ h_{2p} \\ ... \\ h_{qp} \end{bmatrix} = w_1 H_{\alpha,1} + w_2 H_{\alpha,2} + .... + w_p H_{\alpha,p}$$

To see the geometric image of the matrix H, for simplicity we consider a three dimensional space for which we have

$$H = \begin{bmatrix} h_{11} & h_{12} & h_{13} \\ h_{21} & h_{22} & h_{23} \\ h_{31} & h_{32} & h_{33} \end{bmatrix}$$

The column vectors in H are a set of three vectors in the three dimensional space of the objects. In Figure 3 we show the image of the vectors in H for attributes $H_i$ (i = 1, 2, 3) and the space of the objects $A_i$ (i = 1, 2, 3).

$$H_1 = \begin{bmatrix} h_{11} \\ h_{21} \\ h_{31} \end{bmatrix} , \ H_2 = \begin{bmatrix} h_{12} \\ h_{22} \\ h_{32} \end{bmatrix} , H_3 = \begin{bmatrix} h_{13} \\ h_{23} \\ h_{33} \end{bmatrix}$$

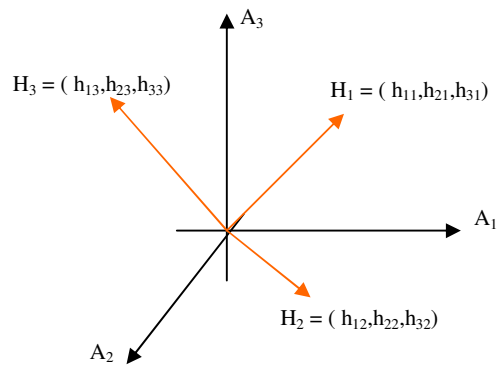

**Figure 3:** Objects $A_k$ and Attributes $H_j$ as Column Vectors in Matrix H

The vector W = $(w_1, w_2, w_3)$ has three components $w_1, w_2, w_3$ on the reference given by three vectors $H_1$, $H_2$, $H_3$ as shown in Figure 4:

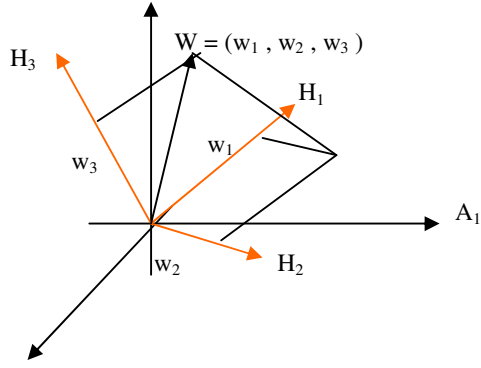

**Figure 4:** Objects $A_k$ and Attributes $H_j$ represented by column vectors in matrix H and the coordinates for the vector W

In classical tensor calculus, the equation becomes

$$y_k = \sum_{j=1}^{p} h_k^j w_j = h_k^j w_j \quad (1.3)$$

When

$$H^T H = \begin{bmatrix} h_{11} & h_{12} & ... & h_{1p} \\ h_{21} & h_{22} & ... & h_{2p} \\ ... & ... & ... & ... \\ h_{q1} & h_{q2} & ... & h_{qp} \end{bmatrix}^T \begin{bmatrix} h_{11} & h_{12} & ... & h_{1p} \\ h_{21} & h_{22} & ... & h_{2p} \\ ... & ... & ... & ... \\ h_{q1} & h_{q2} & ... & h_{qp} \end{bmatrix} =$$

$$\begin{bmatrix} 1 & 0 & ... & 0 \\ 0 & 1 & ... & 0 \\ ... & ... & ... & ... \\ 0 & 0 & ... & 1 \end{bmatrix} = \delta_{h,k}$$

The set of vectors or samples $H_k$ are one orthogonal to the others.

### 3.1  projection operator Q and weights w

Now since Y = H W, and because H is a rectangular matrix, for the pseudo inverse matrix we have

$$W = (H^T H)^{-1} H^T Y$$

$$HW = H(H^T H)^{-1} H^T Y$$

$$For$$

$$Y = HW$$

$$HW = H(H^T H)^{-1} H^T HW = HW = Y$$

Where

$$Q = H \, ( \, H^T \, H \, )^{-1} \, H^T$$

Is the projection operator previously defined. For $g = H^T \, H$, we can write by index notation in tensor calculus

$$W = x^h = g^{-1} x_k = (H^T H)^{-1} H^T Y$$

Where $W = x^h, H^T X = x_k$ in tensor calculus g is the metric tensor and

$$g = H^T H = \sum_k h_{kj} h_{ki} = h_j^k h_{ki} \ (2.3)$$

The covariant components of X are:

$$x_k = H^T X = h_k^j A_j \ (3.3)$$

Where $X_j$ are the components of X in the space of the object. The contravariant components of X are:

$$x^i = g^{-1} x_j = g^{i,j} x_j = W \ (4.3)$$

We note that *when H is a square matrix* we obtain:

$$Y = HW = h_{j,i} x^i = A_j = X$$

When H is a rectangular matrix with q > p, we have

$$QX = HW = H(H^T H)^{-1} H^T X = h_{j,i} x^i = y_j$$

With the property:

$$Q^2 X = H(H^T H)^{-1} H^T H(H^T H)^{-1} H^T X = H(H^T H)^{-1} H^T X = QX$$

Q is a projection operator. In figure 5 we show the geometric image of the projection operator for three dimensional object space and two dimensional attribute space.

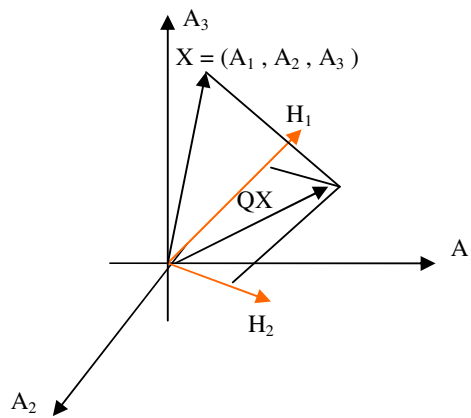

**Figure 5:** Example of the Projection Operator Q

As an example:

$$H = \begin{bmatrix} \cos(\alpha) & \cos(\beta) \\ \sin(\alpha) & \sin(\beta) \end{bmatrix}, g = H^T H = \begin{bmatrix} 1 & \cos(\alpha - \beta) \\ \cos(\alpha - \beta) & 1 \end{bmatrix}$$

Given the vectors $X = \begin{bmatrix} A_1 \\ A_2 \end{bmatrix}$ we have:

$$x_k = \begin{bmatrix} x_1 \\ x_2 \end{bmatrix} = H^T X = \begin{bmatrix} \cos(\alpha) & \sin(\alpha) \\ \cos(\beta) & \sin(\beta) \end{bmatrix} \begin{bmatrix} A_1 \\ A_2 \end{bmatrix} = \begin{bmatrix} A_1 \cos(\alpha) + A_2 \sin(\alpha) \\ A_1 \cos(\beta) + A_2 \sin(\beta) \end{bmatrix}$$

$$x^h = g^{-1} x_k = \begin{bmatrix} \dfrac{(A_1 \cos(\alpha) + A_2 \sin(\alpha)) - \cos(\alpha - \beta)(A_1 \cos(\beta) + A_2 \sin(\beta))}{\sin^2(\alpha - \beta)} \\ \dfrac{(A_1 \cos(\beta) + A_2 \sin(\beta)) - \cos(\alpha - \beta)(A_1 \cos(\alpha) + A_2 \sin(\alpha))}{\sin^2(\alpha - \beta)} \end{bmatrix}$$

The two components of X can be represented by the graph in figure 6:

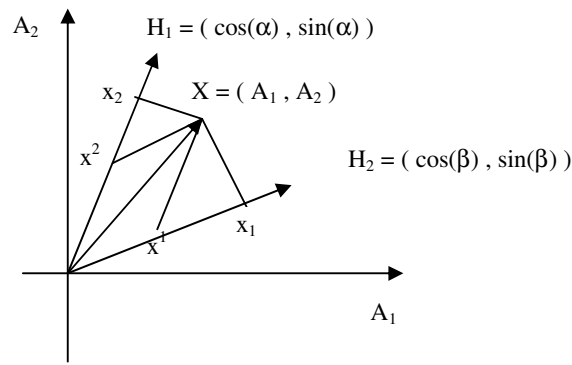

**Figure 6:** General coordinates $H_1$, $H_2$, Object coordinates $A_1$, $A_2$.
Components of X, Covariant components $x_k$ and contravariant components $x^k$ Q

Another example of the projection operator is given by the change of coordinates

$$\begin{cases} \bar{z}_1 = \bar{z}_1(z_1, z_2, \ldots, z_n) \\ \bar{z}2 = \bar{z}2(z_1, z_2, \ldots, z_n) \\ \quad \ldots \\ \bar{z}n = \bar{z}n(z_1, z_2, \ldots, z_n) \end{cases}$$

Where the matrix U s:

$$U = \begin{bmatrix} \dfrac{\partial z_1}{\partial \bar{z}_1} & \dfrac{\partial z_2}{\partial \bar{z}_1} & \ldots & \dfrac{\partial z_n}{\partial \bar{z}_1} \\ \dfrac{\partial z_1}{\partial \bar{z}2} & \dfrac{\partial z_2}{\partial \bar{z}2} & \ldots & \dfrac{\partial z_n}{\partial \bar{z}2} \\ \ldots & \ldots & \ldots & \ldots \\ \dfrac{\partial z_1}{\partial \bar{z}n} & \dfrac{\partial z_2}{\partial \bar{z}n} & \ldots & \dfrac{\partial z_n}{\partial \bar{z}n} \end{bmatrix}$$

And

$$U^{-1} = \begin{bmatrix} \dfrac{\partial \overline{z}_1}{\partial z_1} & \dfrac{\partial \overline{z}_2}{\partial z_1} & \cdots & \dfrac{\partial \overline{z}_n}{\partial z_1} \\[2mm] \dfrac{\partial \overline{z}_1}{\partial z_2} & \dfrac{\partial \overline{z}_2}{\partial z_2} & \cdots & \dfrac{\partial \overline{z}_n}{\partial z_2} \\[2mm] \cdots & \cdots & \cdots & \cdots \\[2mm] \dfrac{\partial \overline{z}_1}{\partial z_n} & \dfrac{\partial \overline{z}_2}{\partial z_n} & \cdots & \dfrac{\partial \overline{z}_n}{\partial z_n} \end{bmatrix}$$

A simple system $X = x_k$ is covariant when it changes in this way:

$$\overline{x}_k = U \ x_j = \sum_j \frac{\partial z^k}{\partial \overline{z}_j} x_j \quad (5.3)$$

Where $x_j$ is the Euclidean geometry and in our theory is located in the object space. A system is contravariant when changes in this way:

$$\overline{x}^k = U^{-1} x^j = \sum_j \frac{\partial \overline{z}^k}{\partial z_j} x^j \quad (6.3)$$

In the projection operator we have

Y = Q X = H W = H $x^k$ = H $g^{-1}$ $x_k$

For the transformation U we have

$$\overline{Y} = H \overline{x}^k = H U^{-1} g^{-1} U x_j = H G^{-1} x_j$$

*where*

$$G = U^{-1} g U$$

So G in the projection operator is equivalent to g. In fact we have that

$$P = \overline{x}^k \overline{x}_k = U^{-1} x^j U x_j = x^j x_j = \sum_j x^j x_j \quad (7.3)$$

Where P is the metric of the space.

# 4.  General Definition of the forces and fluxes

In the projection operator $Q = A( B^T A )^{-1} B^T$ where A are samples in the space of the causes as objects and B are samples in the space of the effects we introduce the purpose X in the space of the causes.

Now we have

$$Q X = A( B^T A )^{-1} B^T X = Y$$

The vector X is a virtual cause of force. In fact the purpose is not an effective cause, but just a project of possible cause that must be translated into the particular context where the virtual cause becomes a real cause.

With A we can also compute the sources of the forces E. Now for the relation:

$$V = \Omega I$$

Where $\Omega$ is the cross matrix representing the couple between the flux J and the force V in the system. We have also:

$$E = A^T V = A^T \Omega \ A J = Z J$$

The matrix Z is the cross matrix for the sources of the force E and the source of the flux J. Given the forces V in all the edges of the network, we compute the sources E in this way:

$$E = A^T V$$

And also the sources of the fluxes:

$$J = Z^{-1} E$$

Now we can compute the invariant form for unitary transformation U or *couple variable between the forces and fluxes:*

$$L = J^T E = J^T Z J = E^T Z^{-1} E$$

We have also that:

$$L = J^T A^T \Omega A J = (A J)^T \Omega A J = I^T \Omega I = I^T V$$

For the unitary transformation U for which $U^T = U^1$. In fact we have:

$$J' = U J, E' = U E$$

and:

$$L' = ( U J )^T U E = J^T U^T U E = J^T E = L$$

Now we can also prove that L under constraints can assume an extreme value. In fact we have:

$$L = J^T Z J + \lambda^T (E - A^T V) = J^T Z J + \lambda^T (E - (A^T \Omega A) J) = J^T Z J + \lambda^T (E - Z J) \ (1.4)$$

Where $E = A^T V$ is the constraint and the vector of $\lambda$ is the Lagrange multiplier. Now for the previous expression we have:

$$\frac{\partial L}{\partial J_p} = \sum_k Z_{p,k}(2J_k - \lambda_k) = 0$$

Now the derivatives are equal to zero for the extreme value where L is assumed to be the maximum or minimum value for the constraint E = Z J. The solution of the previous system gives the solution

$$\lambda_k = 2J_k \quad (2.4)$$

When we substitute (2.4) into (1.4 ) we have

$$L = J^T Z J + 2J^T(E - Z J) = J^T Z J + 2J^T E - 2J^T Z J = 2J^T E - J^T Z J \quad (3.4)$$

Now when we compute the derivative we have

$$\frac{\partial L}{\partial J_p} = 2E - 2Z J = 0, E = Z J \quad (4.4)$$

When we substitute (4.4) into (3.4) we have

$$L = 2J^T Z J - J^T Z J = J^T Z J$$

In conclusion, the sources of currents are located in the system so as to satisfy the minimum condition for L. We remark also that for the previous computation we have:

$$V' = \Omega A J = \Omega A (A^T \Omega A)^{-1} A^T V = Q V$$

and

$$A^T Q V = A^T V$$

In fact we have

$$A^T Q V = A^T \Omega A (A^T \Omega A)^{-1} A^T V = A^T V$$

The product of V and QV with $A^T$ is invariant. For the operator Q we have

$$Q^2 = \Omega A (A^T \Omega A)^{-1} A^T \Omega A (A^T \Omega A)^{-1} A^T = \Omega A (A^T \Omega A)^{-1} A^T = Q$$

So Q is a projection operator.

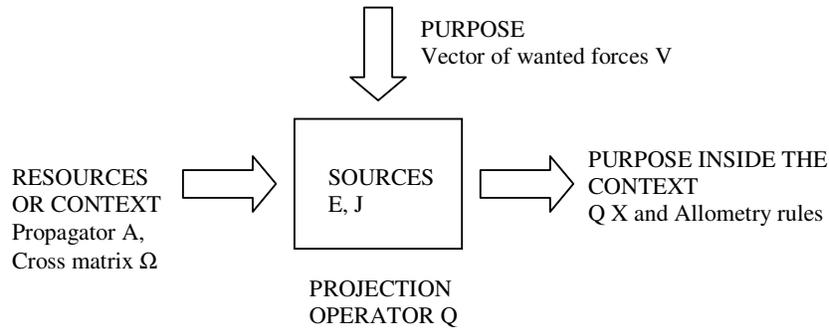

**Figure 7:** Graphic image of the projection operator and Morphogenic System

For the definition of the projection operator we have:

$$L = J^T E = ((A^T \Omega A)^{-1} A^T V)^T A^T V = V(A(A^T \Omega A)^{-1} A^T V) = (V)^T \Omega^{-1}(QV)$$

And we have also that:

$$A^T QV = A^T \Omega A(A^T \Omega A)^{-1} A^T V = A^T V = E$$

The sources of forces E are the invariants for the projection operator. For a congruent system, two systems are congruent when we have:

$$J' = U J, E' = (Z U) J$$

Where U is the unitary transformation for which $U^T U = 1$. Now we have

$$L' = J' E' = (U J)^T Z (U J) = J^T U^T Z U J = J^T Z' J$$

We remark that L can be invariant for the transformation of the current J into UJ when we transform Z in this way

$$Z' = U Z U^T$$

In fact we have

$$L' = (U J)^T U Z U^T (U J) = J^T Z J = L$$

Given the space of fluxes J, where any component of space is the value of the current, we can represent Z as the metric of space and L as a distance in the fluxes' space. Now given the vector V, the force E is the covariant component of V and the flux J is the contravariant component of V in non Euclidean geometry. Graphically we have:

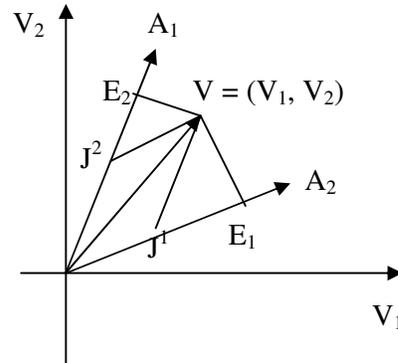

**Figure 8:** Given the matrix A = (A₁, A₂) and force V = (V₁, V₂),
the sources of forces E₁, E₂ and the sources of fluxes J₁, J₂

The Projection operator and geometric image is shown in Figure 9:

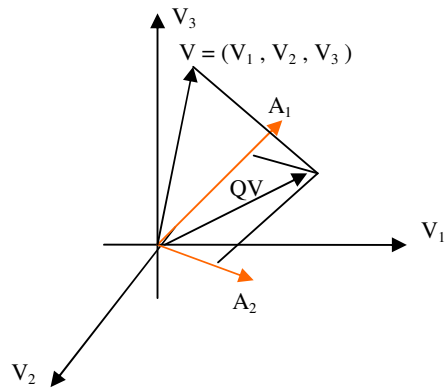

**Figure 9:** Geometric image of the projection operator Q

## 4.1 Invariants

The space of objects includes the vectors of the attributes. In this case, any vector X is projected in the subspace of the object defined by vectors of attributes. Now given QX = $y_k$, we can compute the scalar:

$$s^2 = (QX)^T (QX) = \sum_k y_k y_k = y^k y_k$$

We note the space of the objects is a Cartesian Space with the Euclidean geometry, so:

$$y^k = y_k$$

In general coordinates the transformation of the components QX in the space of objects into components of attributes are

$$\overline{y}_k = H^T QX = h_k^{\ j} y_j$$

and

$$\overline{y}^k = (H^T H)^{-1} H^T QX = g^{-1} h_k^{\ j} y_j$$

For the definition of the projection operator, we have

$$\overline{y}_k = H^T H (H^T H)^{-1} H^T X = H^T X = h_k^j x_j = \overline{x}_k$$

and

$$\overline{y}^k = (H^T H)^{-1} H^T H (H^T H)^{-1} H^T X = (H^T H)^{-1} H^T X = g^{-1} h_k^j x_j = \overline{x}^k$$

Now the scalar can be computed with new components in this way:

$$s^2 = \overline{y}^k \, \overline{y}_k = ((H^T H)^{-1} H^T X)^T \, H^T X = X^T H (H^T H)^{-1} (H^T H)(H^T H)^{-1} H^T X =$$

$$X^T (H (H^T H)^{-1} H^T) H (H^T H)^{-1} H^T X = X^T Q^T Q X = (Q X)^T (Q X) = y^k \, y_k$$

Now we want to solve the equation:

$$Q X = X$$

When H is a quadratic matrix we have:

$$QX = H (H^T H)^{-1} H^T X = H (H)^{-1} (H^T)^{-1} H^T X = X$$

When $H = \begin{bmatrix} A \\ B \end{bmatrix}$, where A is a quadratic matrix with the p dimension, we can write B = C A, where C = B $A^{-1}$. Now when $X = \begin{bmatrix} X_A \\ X_B \end{bmatrix}$, where $X_A$ is a vector with p dimensions and $X_B$ = C $X_A$, we can express that Q X = X. In fact we have:

$$QX = \begin{bmatrix} A \\ CA \end{bmatrix} (\begin{bmatrix} A \\ CA \end{bmatrix}^T \begin{bmatrix} A \\ CA \end{bmatrix})^{-1} \begin{bmatrix} A \\ CA \end{bmatrix}^T \begin{bmatrix} X_A \\ CX_A \end{bmatrix} = \begin{bmatrix} A \\ CA \end{bmatrix} (\begin{bmatrix} A & CA \end{bmatrix} \begin{bmatrix} A \\ CA \end{bmatrix})^{-1} \begin{bmatrix} A \\ CA \end{bmatrix}^T \begin{bmatrix} X_A \\ CX_A \end{bmatrix} =$$

$$\begin{bmatrix} A \\ CA \end{bmatrix} (A^T A + (CA)^T (CA))^{-1} \begin{bmatrix} A \\ CA \end{bmatrix}^T \begin{bmatrix} X_A \\ CX_A \end{bmatrix} = \begin{bmatrix} A \\ CA \end{bmatrix} (A^T A + A^T C^T CA))^{-1} \begin{bmatrix} A \\ CA \end{bmatrix}^T \begin{bmatrix} X_A \\ CX_A \end{bmatrix} =$$

$$\begin{bmatrix} A \\ CA \end{bmatrix} (A^T (A + C^T CA))^{-1} \begin{bmatrix} A \\ CA \end{bmatrix}^T \begin{bmatrix} X_A \\ CX_A \end{bmatrix} = \begin{bmatrix} A \\ CA \end{bmatrix} (A^T (I + C^T C)A)^{-1} \begin{bmatrix} A \\ CA \end{bmatrix}^T \begin{bmatrix} X_A \\ CX_A \end{bmatrix} =$$

$$\begin{bmatrix} A \\ CA \end{bmatrix} A^{-1} (I + C^T C)^{-1} (A^T)^{-1} \begin{bmatrix} A \\ CA \end{bmatrix}^T \begin{bmatrix} X_A \\ CX_A \end{bmatrix} = \begin{bmatrix} I \\ C \end{bmatrix} (I + C^T C)^{-1} \begin{bmatrix} I & C^T \end{bmatrix} \begin{bmatrix} X_A \\ CX_A \end{bmatrix} =$$

$$\begin{bmatrix} I \\ C \end{bmatrix} (I + C^T C)^{-1} (I + C^T C) X_A = \begin{bmatrix} I \\ C \end{bmatrix} X_A = \begin{bmatrix} X_A \\ CX_A \end{bmatrix} = X$$

When QX = X, we have:

$$s^2 = \overline{y}^k \, \overline{y}_k = x^k x_k$$

Because x is a vector in the Euclidean Space of objects, $x^k = x_k$ and:

$$s^2 = \overline{y}^k \, \overline{y}_k = x^k x_k = \sum_k x_k^2$$

In this case $s^2$ is invariant for any context H for which Q X = X. As an example:

For $H = \begin{bmatrix} 1 & 0 \\ 1 & -1 \\ 0 & 1 \end{bmatrix}$ we have $A = \begin{bmatrix} 1 & 0 \\ 1 & -1 \end{bmatrix}$, $B = \begin{bmatrix} 0 & 1 \end{bmatrix}$

So we have $C = BA^{-1} = \begin{bmatrix} 1 & -1 \end{bmatrix}$. Now for $X = \begin{bmatrix} x_1 \\ x_2 \\ C\begin{bmatrix} x_1 \\ x_2 \end{bmatrix} \end{bmatrix} = \begin{bmatrix} x_1 \\ x_2 \\ x_1 - x_2 \end{bmatrix}$

We have:

$$Q \begin{bmatrix} x_1 \\ x_2 \\ x_1 - x_2 \end{bmatrix} = \begin{bmatrix} x_1 \\ x_2 \\ x_1 - x_2 \end{bmatrix}$$

Now for $X = \begin{bmatrix} X_A \\ X_B \end{bmatrix}$ we have:

$$QX = \begin{bmatrix} I \\ C \end{bmatrix} (I + C^T C)^{-1} \begin{bmatrix} I & C^T \end{bmatrix} \begin{bmatrix} X_A \\ X_B \end{bmatrix} = \begin{bmatrix} I \\ C \end{bmatrix} (I + C^T C)^{-1} (X_A + C^T X_B) =$$

$$\begin{bmatrix} I \\ C \end{bmatrix} (I + C^T C)^{-1} (X_A + C^T (CX_A + D)) = \begin{bmatrix} I \\ C \end{bmatrix} X_A + \begin{bmatrix} I \\ C \end{bmatrix} (I + C^T C)^{-1} C^T D =$$

$$\begin{bmatrix} X_A + (I + C^T C)^{-1} C^T D \\ C(X_A + (I + C^T C)^{-1} C^T D) \end{bmatrix} = \begin{bmatrix} X_A + E \\ C(X_A + E) \end{bmatrix}$$

In conclusion we have that:

$$s^2 = \overline{y}^k \ \overline{y}_k = \overline{x}^k \ \overline{x}_k = \ \overline{y}^k \ \overline{y}_k \quad (5.4)$$

The scalar $s^2$ is invariant when we move from the space of the objects to the space of the attributes. The expression (5.4) can be written in this way:

$$s^2 = \overline{y}^k \ \overline{y}_k = \overline{y}^k \ g \ \overline{y}^k = \overline{x}^k \ g \ \overline{x}^k = \overline{y}_k \ g^{-1} \ \overline{y}_k = \overline{x}_k \ g^{-1} \ \overline{x}_k \quad (6.4)$$

## 4.2 The Projection Operator and Lagrangian Function Minimum Condition

We separate the projection operator in two parts one is the source of the force E and the other is the source of the flux J. The two parts are:

$$J = x^h = (B^T A)^{-1} B^T X \quad and \quad E = B^T X = x_k$$

Now the Lagrangian form is written as:

$$L = x^h g x^h = J^T g J + \lambda(E - B^T Q X) = J^T g J + \lambda(E - B^T A J) = J^T g J + \lambda^T (E - g J)$$

The minimum condition is:

$$\frac{dL}{dJ} = 2 g J - \lambda g = 0$$

The solution is $\lambda = \frac{J}{2}$ when we substitute in L we have:

$$L = J^T g J + \frac{1}{2} J^T (E - g J) = \frac{1}{2} J^T g J + \frac{1}{2} J^T E$$

Now for the definition of E and J we have

$$L = \frac{1}{2} J^T g J + \frac{1}{2} J^T E = \frac{1}{2} J^T g J + \frac{1}{2} J^T g J = J^T g J = J^T E = x^h x_h$$

So the for invariant $L = J^T g J$, one assumes the minimum value under the constraint $E = g J$.

## 5. The Quantum Connection

The essential trait of Quantum Physics is non-locality, which could naturally perform the hypercomputation's feature: exploring "many worlds" in finite time by means of superposition and entanglement (Copeland & Proudfoot, 1999; Ord, 2002; Licata 2007; Hagar & Korolev, 2007; Syropoulos, 2008).

In order to understand the relation between non-locality and the possibility to realize oracles it is useful to make reference to quantum potential and the notion associated to active information introduced by David Bohm and the Birbeck College group (Bohm and Hiley, 2005; Callaghan, Hiley and Maroney, 2000; Hiley and Maroney, 1999, 2000; Hiley, 2002). Let us here remember that differently from what is usually said Bohm's "non-mechanics" (Hiley, 2000) is formally equivalent to Standard QM, but has the merit to include non-locality *ab initio* rather than to come upon it as an *a posteriori* statistical "mysterious weirdness". As it is known, the Quantum Potential (QP) derives from decomposing the Schrödinger Equation into real part and imaginary part by using the polar expression for the wave function $\Psi(r,t) = R(r,t) \exp[iS(r,t)/\hbar]$.

So, for the Quantum Potential, we have:

$$Q(r,t) = -\frac{\hbar^2}{2m} \frac{\nabla^2 R(r,t)}{R(r,t)} \quad (1.5)$$

The Quantum Potential has a contextual nature, i.e. it provides global information about the process and its environment by individuating an infinite set of phase paths. We underline that such reading is perfectly coherent with the formulation of Feynman's path integrals. Active information described by (1.5) is deeply different from classical one: it is, in fact, intrinsically not-Shannon computable: if it were not so, Bell's inequalities on the impossibility of a QM with local hidden variables would be violated.

All that has a deep physical significance as for the dynamics of quantum information. The quantum potential indicates a source of active information internal to the system and differently accessible according to the operations of preparation and state selection, environment and measurement. The quantum system's active information is defined by an infinite uncountable set of phase paths and its very nature is non-local. In such configurations as "quantum gates", based upon a generalization of Turing scheme, the constrains of reversibility and unitarity limit the possibility to detect quantum information just to the outputs of superposition states; and yet nothing prevents our thinking of a different approach to the system's geometry which, within peculiar experimental arrangements, can get qualitatively different answers and endowed with oracular skills, so turning into resource all non-locality features, even those which are traditionally regarded as a limit within the classical scheme, such as de-coherence, dissipation and probabilistic responses.

An essential key for the relations between quantum potential, system's geometry and information is provided by the Fisher information (Carroll, 2006).

The Fisher information can be interpreted as the information an observable random variable X carries about a not-observable parameter θ which the probability distribution X depends on. It is clear that such statistical measurement has aroused interest in relation to the study of the distributions of the observables of a quantum system.

As for the physical meaning of the Fisher information, instead, there are very controversial viewpoints.

Roy Frieden's programme (Frieden, 2004) to derive Physics' fundamental equations from an extreme physical Fisher information principle as optimization (or saturation) of the observer/observed relationship has been widely criticized because of its vagueness. Actually, although Frieden's position is epistemologically correct for a good experimental physicist as he is, the principle itself is too less constraining, so that the significant physical features of the systems under observation have to be introduced in order to make it really effective (Streater, 2007). Thus, Frieden's programme looks more like a request for coherence between formal structures and distributions of observables than an out-and-out "fundamental principle".

The studies where an attempt is made to connect Fisher information with the specific structural aspects of Quantum Mechanics and to consider it as a statistical indicator of the relationships between classical and quantum information are more interesting (Hall, 2000; Luo Shun-Long, 2006; Luati, 2004). In spite of the "interpretative dilemmas", Quantum Mechanics shows the highest operational nature of any other physic theory, and it is thus greatly interesting that the "thin" statistical distribution of a quantum system can be derived from the quantum potential. In particular, it has been shown (Hradil & Rehacek, 2004) that the lower bound for Fisher information, Cramer and Rao inequality, in a quantum process fixes the complementarity of the variables according to Heisenberg principle; for the variables of position x and momentum p, by putting Fisher information as F, we have:

$$(\Delta p)^2 \geq \frac{1}{F} \geq \frac{\hbar^2}{4(\Delta x)^2}$$

Our aim is to show how Fisher information plays the role of a natural tile to build a metric able to connect the system's statistical outcomes and its global geometry. In order to do it, in the next paragraphs we will examine the Schrödinger equation and analyse the double slit classical experiment.

### 5.1 The Structure of Schrödinger Equation and Fisher information

Given the classical Hamilton Jacobi equation

$$Q = \frac{\partial S}{\partial t} + \frac{1}{2m}(\frac{\partial S}{\partial x})^2$$

Where

$$mv = \frac{\partial S}{\partial x} = p \qquad ,$$

the average value A of Q is given by the expression

$$A = -\int_{t_0}^{t_1}\int_{-\infty}^{\infty} (\frac{\partial S}{\partial t} + \frac{1}{2m}(\frac{\partial S}{\partial x})^2) P^i(x,t,\theta)dxdt = -\int_{t_0}^{t_1}\int_{-\infty}^{\infty} f(\frac{\partial S}{\partial t},\frac{\partial S}{\partial x})dxdt$$

Now we assume that the average value of A has variation equal to zero, so we have δ A = 0. For the Euler Lagrange equation we obtain

$$\frac{\partial}{\partial t}\left(\frac{\partial f}{\partial \frac{\partial S}{\partial t}}\right) + \frac{\partial}{\partial x}\left(\frac{\partial f}{\partial \frac{\partial S}{\partial x}}\right) = 0$$

For which we have

$$\frac{\partial P^i}{\partial t} + \frac{\partial}{\partial x}\left(P^i \frac{1}{m}\frac{\partial S}{\partial x}\right) = 0$$

That is the continuous equation. Now for

$$A_{Diffusion} = -\int\limits_{t_0}^{t_1}\int\limits_{-\infty}^{\infty} P(t,\theta(x))\left\{\frac{\partial S}{\partial t} + \frac{1}{2m}\left[\left(\frac{\partial S}{\partial x}\right)^2 - \frac{1}{P^2}\left(\frac{\partial P(t,\theta(x))}{\partial \theta}\right)^2\right]\right\}dxdt =$$

$$-\int\limits_{t_0}^{t_1}\int\limits_{-\infty}^{\infty} P(t,\theta(x))\left\{\frac{\partial S}{\partial t} + \frac{1}{2m}\left[\left(\frac{\partial S}{\partial x}\right)^2 - \frac{1}{P^2}\left(\frac{\partial P(t,\theta(x))}{\partial x}\frac{\partial x}{\partial \theta}\right)^2\right]\right\}dxdt =$$

$$-\int\limits_{t_0}^{t_1}\int\limits_{-\infty}^{\infty} P(t,\theta(x))\left\{\frac{\partial S}{\partial t} + \frac{1}{2m}\left[\left(\frac{\partial S}{\partial x}\right)^2 - \frac{D^2}{P^2}\left(\frac{\partial P(t,\theta(x))}{\partial x}\right)^2\right]\right\}dxdt$$

Where :

$$F = -\int\limits_{-\infty}^{\infty} P(t,\theta(x))\frac{1}{P^2}\left(\frac{\partial P(t,\theta(x))}{\partial \theta}\right)^2 dx = -\int\limits_{-\infty}^{\infty}\frac{1}{P}\left(\frac{\partial P(t,\theta(x))}{\partial \theta}\right)^2 dx$$

is the Fisher information. Now we have

$$A_{Diffusion} = -\int\limits_{t_0}^{t_1}\int\limits_{-\infty}^{\infty} f\left(P, \frac{\partial P}{\partial x}\right)dxdt$$

And by fixed end-point variation δ A = 0 we obtain

$$\frac{\partial f}{\partial P} - \frac{\partial}{\partial x}\left(\frac{\partial f}{\partial \frac{\partial P}{\partial x}}\right) = 0$$

and

$$\frac{\partial S}{\partial t} + \frac{1}{2m}\left(\frac{\partial S}{\partial x}\right)^2 + Q = \frac{\partial S}{\partial t} + E + Q = 0$$

Where

$$Q = -\frac{D^2}{4}\left(\frac{1}{P^2}\left(\frac{\partial P}{\partial x}\right)^2 - \frac{2}{P}\left(\frac{\partial^2 P}{\partial x^2}\right)\right) = \frac{D^2}{4\sqrt{P}}\frac{\partial\sqrt{P}}{\partial x},$$

where Q defines the system's quantum potential.
We remember that given the Schrödinger equation

$$ih\frac{\partial \psi}{\partial t} = -(\frac{h^2}{2m})\nabla^2\psi + V(\vec{r})\psi$$

For

$$\psi = R\exp(i\frac{S}{h})$$

We have

$$\frac{dP}{dt} = \frac{\partial P}{\partial t} + \frac{\partial P}{\partial x}v_x + \frac{\partial P}{\partial y}v_y + \frac{\partial P}{\partial y}v_y = \frac{\partial P}{\partial t} + \nabla(P\frac{\nabla S}{m}) = 0$$

$$\frac{\partial S}{\partial t} + \frac{(\nabla S)^2}{2m} + V(r) - \frac{h^2}{2m}\frac{\nabla^2 R}{R} = \frac{\partial S}{\partial t} + \frac{(\nabla S)^2}{2m} + V(r) - \frac{h^2}{2m}\frac{\nabla^2\sqrt{P}}{\sqrt{P}} = 0$$

Now this particular combination of the probability P and S

$$\rho = \sqrt{P}\exp(\frac{S}{D})$$

$$\rho^* = \sqrt{P}\exp(-\frac{S}{D})$$

Given a function $\rho$ and $\rho^*$ that are solution of the diffusion equations

$$\frac{\partial \rho}{\partial t} = \frac{D}{2m}\frac{\partial^2 \rho}{\partial x^2}$$

$$\frac{\partial \rho^*}{\partial t} = -\frac{D}{2m}\frac{\partial^2 \rho^*}{\partial x^2}$$

We point out that we have followed here the traditional Nelson's diffusive notation (Nelson, 2001; Reginatto & Lengyel, 1999), but in no way we mean to put forward any further hypothesis on vacuum. Both the coefficient D and the $A_{diffusion}$ can be more formally regarded as the measurement of the densities of phase trajectories, directly connected to the density matrix as well as the system's entropy (Aharonov & Anandan, 1998).

The Fisher information for a quantum system is directly connected to quantum potential, and characterizes its entropy in terms of distribution of phase trajectories. That provides the concept of active information with a very definite physical meaning in relation to the global structure of the non-local field, and we will show that by defining an opportune metric it is the tile to build the relation between geometry and quantum information.

## 5.2. Quantum mechanics interference and geometry

As Feynman shrewdly pointed out, the double slit experiment contains all the essential features of quantum physics (Feynman, 1970). Thus it is the ideal starting point to introduce the connection between quantum information and geometric covariance.

In this paradigmatic experiment we have interference between two probability densities $h_1$ and $h_2$:

$$|h\rangle = |h_1\rangle + |h_2\rangle =$$

$$A_1(r,t)e^{iS_1(r,t)} + A_2(r,t)e^{iS_2(r,t)} =$$

$$A_1(r,t)\cos(S_1(r,t)) + A_2(r,t)\cos(S_2(r,t)) + i\,[A_1(r,t)\sin(S_1(r,t)) + A_2(r,t)\sin(S_2(r,t))]$$

Where A is the amplitude and S is the phase at specific location r in specific time t. We know that the probability of the interference is

$$P = \langle h | h \rangle =$$

$$\langle h_1 + h_2 | h_1 + h_2 \rangle =$$

$$\langle h_1 | h_1 \rangle + \langle h_2 | h_2 \rangle + 2\langle h_1 | h_2 \rangle$$

because the probability densities $|h_1\rangle, |h_2\rangle$ are complex numbers we have

$$P = \langle h | h \rangle = \;(A_1(r,t)\cos(S_1(r,t)) + A_2(r,t)\cos(S_2(r,t)) + i\,[A_1(r,t)\sin(S_1(r,t)) + A_2(r,t)\sin(S_2(r,t))])$$

$$(A_1(r,t)\cos(S_1(r,t)) + A_2(r,t)\cos(S_2(r,t)) - i\,[A_1(r,t)\sin(S_1(r,t)) + A_2(r,t)\sin(S_2(r,t))]) =$$

$$(A_1(r,t)\cos(S_1(r,t)) + A_2(r,t)\cos(S_2(r,t)))^2 + (A_1(r,t)\sin(S_1(r,t)) + A_2(r,t)\sin(S_2(r,t)))^2 =$$

$$A_1(r,t)^2 + A_2(r,t)^2 + 2A_1(r,t)A_2(r,t)[\cos(S_1(r,t))\cos(S_2(r,t)) + \sin(S_1(r,t))\sin(S_2(r,t))] =$$

$$A_1(r,t)^2 + A_2(r,t)^2 + 2A_1(r,t)A_2(r,t)\cos(S_1(r,t) - S_2(r,t)) =$$

$$I_1(r,t) + I_2(r,t) + 2\sqrt{I_1(r,t)I_2(r,t)}\cos(S_1(r,t) - S_2(r,t))$$

Where $I_1(x)$ and $I_2(x)$ are the intensity of the beam of electrons
Because we have

$$P = P_1 + P_2 + 2\langle h_1 | h_2 \rangle = I_1(x) + I_2(x) + 2\sqrt{I_1(x)I_2(x)}\cos(\alpha - \beta)$$

$$\alpha = S_1(r,t)$$

$$\beta = S_2(r,t)$$

Let's note that quantum probability P is a non-additive probability. In figure 10 we show another interference experiment:

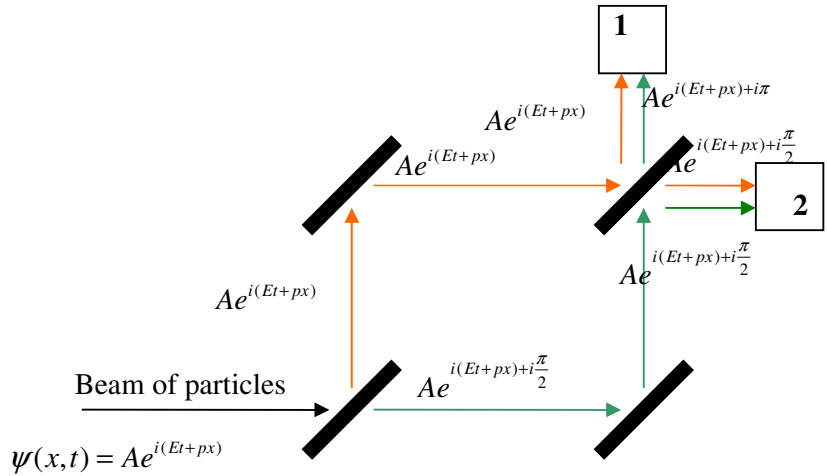

Figure 10 in the first detector we have a destructive interference in the second detector we have the constructive interference.

The first crystal in figure 10 the black line, splits a beam in two beams of particle one with phase zero , red color , the other beam with phase $\frac{\pi}{2}$. The final crystal changes the green beam again of $\frac{\pi}{2}$, so the final green beam up is a phase of $\pi$ and the read beam up has phase zero. The final green beam on the right does not change the phase, the red beam on the right changes the phase by zero to $\frac{\pi}{2}$. The detector 1 in up receives two beams, red and green, with opposite phase so the interference gives zero intensity for the beam, the interference is destructive. The detector 2 on the right receives two beams with the same phase so we have the maximum intensity of the interference, the interference is constructive.

In figure 10 we have both constructive and destructive interference. From the previous consideration we have that the geometry of the system of the first detector is different from the second detector one.

In the first detector we have the metric P(x) defined in this way:

$$P(x) = I_1^2 + I_2^2 + 2\cos(0 - \pi)I_1 I_2 = I_1^2 + I_2^2 - 2I_1 I_2 = (I_1 - I_2)^2 = 0 \qquad (2.5)$$

Because in the second detector we have the metric

$$P(x) = I_1^2 + I_2^2 + 2\cos(\frac{\pi}{2} - \frac{\pi}{2})I_1 I_2 = I_1^2 + I_2^2 + 2I_1 I_2 = (I_1 + I_2)^2 = 4A^2 \qquad (3.5)$$

The intensity of the interference is the maximum value that is higher than the intensity $A^2$ of the original beam, because it depends on the non-local influence of the geometry in the second detector.

When the two beams are independent one from the other, we have the traditional expression for the distance:

$$P(x) = \psi_1^2 + \psi_2^2 = I_1(x) + I_2(x) = I_1 + I_2 \ ,$$

where no interference is possible.

When the two paths are superposed the two beams are considered as a single beam. The probability P(x) is given by the contribute of three terms

$$P = P_1 + P_2 + P_3$$
$$P_1 = I_1$$
$$P_2 = I_2 \qquad\qquad (4.5)$$
$$P_3 = \sqrt{I_1 I_2}\cos(\alpha - \beta)$$

The third probability is the entanglement between the particles. In a more general case we have

$$P(\psi_1, \psi_2, ...., \psi_n) = m(\langle \psi | \psi \rangle) \qquad (5.5)$$

where

$$\psi = h_1 + h_2 + ..... + h_n$$
$$h_1 = \alpha_1 + i\beta_1 \ , \ h_2 = \alpha_2 + i\beta_2, ....., h_n = \alpha_n + i\beta_n$$

and

$$P(x) = \langle \psi | \psi \rangle = \langle h_1 + h_2 + .... + h_n | h_1 + h_2 + .... + h_n \rangle =$$
$$\langle h_1 | h_1 \rangle + \langle h_2 | h_2 \rangle + ..... + \langle h_n | h_n \rangle + 2\langle h_1 | h_2 \rangle + ..... + 2\langle h_{n-1} | h_n \rangle =$$
$$I_1 + I_2 + ... + I_n + 2\sqrt{I_1 I_2}\cos(\alpha_1 - \alpha_2) + 2\sqrt{I_1 I_3}\cos(\alpha_1 - \alpha_3) + ... + 2\sqrt{I_n I_{n-1}}\cos(\alpha_{n-1} - \alpha_n) =$$
$$\sum_{i,j} g_{i,j} \xi^i(x)\xi^j(x) = s^2$$

where

$$\xi^i = \sqrt{I^i} \quad \xi^j = \sqrt{I^j}$$

and

$$g = \begin{bmatrix} 1 & \cos(\alpha_1 - \alpha_2) & ... & \cos(\alpha_1 - \alpha_2) \\ \cos(\alpha_2 - \alpha_1) & 1 & ... & \cos(\alpha_2 - \alpha_1) \\ ... & ... & ... & ... \\ \cos(\alpha_n - \alpha_1) & \cos(\alpha_n - \alpha_2) & ... & 1 \end{bmatrix}$$

$$g = \begin{bmatrix} 1 & \cos(\alpha_1 - \alpha_2) & ... & \cos(\alpha_1 - \alpha_n) \\ \cos(\alpha_2 - \alpha_1) & 1 & ... & \cos(\alpha_2 - \alpha_n) \\ ... & ... & ... & ... \\ \cos(\alpha_n - \alpha_1) & \cos(\alpha_n - \alpha_2) & ... & 1 \end{bmatrix} = \begin{bmatrix} g_{11} & g_{12} & ... & g_{1n} \\ g_{n1} & g_{22} & ... & g_{2n} \\ ... & ... & ... & ... \\ g_{n1} & g_{n2} & ... & g_{nn} \end{bmatrix} \qquad (6.5)$$

With the metric (6.5) we note that the square distance $s^2$ between the end-points of two vectors $\xi^i$ and $\eta^i$ is

$$s^2(\theta_k) = \sum_{i,j} g_{i,j} (\xi^i(\theta_k) - \eta^i(\theta_k))(\xi^j(\theta_k) - \eta^j(\theta_k)) \qquad (7.5)$$

When

$$\eta^i = \xi^i + \frac{\partial \xi^i}{\partial \theta_k} \partial \theta^k$$

We have

$$ds^2 = \sum_{i,j} g_{i,j} \left( \frac{\partial \xi^i}{\partial \theta_k} \partial \theta^k \right)\left( \frac{\partial \xi^j}{\partial \theta_h} \partial \theta^h \right) =$$

$$\sum_{i,j} g_{i,j} \left( \frac{\partial \xi^i}{\partial \theta_1} \partial \theta_1 + \frac{\partial \xi^i}{\partial \theta_2} \partial \theta_2 + \dots + \frac{\partial \xi^i}{\partial \theta_q} \partial \theta_q \right)\left( \frac{\partial \xi^j}{\partial \theta_1} \partial \theta_2 + \frac{\partial \xi^j}{\partial \theta_2} \partial \theta_2 + \dots + \frac{\partial \xi^j}{\partial \theta_q} \partial \theta_q \right) = \qquad (8.5)$$

$$\sum_{h,k} \left( \sum_{i,j} g_{i,j} \frac{\partial \xi^i}{\partial \theta_h} \frac{\partial \xi^j}{\partial \theta_k} \right)\partial \theta_h \partial \theta_k = \sum_{h,k} G_{h,k} \partial \theta_h \partial \theta_k = G_{h,k} \partial \theta^h \partial \theta^k$$

So we have

$$G = G_{h,k} = A^T g A = \begin{bmatrix} \frac{\partial \xi^1}{\partial \theta_1} & \frac{\partial \xi^1}{\partial \theta_2} & \dots & \frac{\partial \xi^1}{\partial \theta_q} \\ \frac{\partial \xi^2}{\partial \theta_1} & \frac{\partial \xi^2}{\partial \theta_2} & \dots & \frac{\partial \xi^2}{\partial \theta_q} \\ \dots & \dots & \dots & \dots \\ \frac{\partial \xi^n}{\partial \theta_1} & \frac{\partial \xi^n}{\partial \theta_2} & \dots & \frac{\partial \xi^n}{\partial \theta_q} \end{bmatrix}^T \begin{bmatrix} g_{1,1} & g_{1,2} & \dots & g_{1,n} \\ g_{2,1} & g_{2,2} & \dots & g_{2,n} \\ \dots & \dots & \dots & \dots \\ g_{n,1} & g_{n,2} & \dots & g_{n,n} \end{bmatrix} \begin{bmatrix} \frac{\partial \xi^1}{\partial \theta_1} & \frac{\partial \xi^1}{\partial \theta_2} & \dots & \frac{\partial \xi^1}{\partial \theta_q} \\ \frac{\partial \xi^2}{\partial \theta_1} & \frac{\partial \xi^2}{\partial \theta_2} & \dots & \frac{\partial \xi^2}{\partial \theta_q} \\ \dots & \dots & \dots & \dots \\ \frac{\partial \xi^n}{\partial \theta_1} & \frac{\partial \xi^n}{\partial \theta_2} & \dots & \frac{\partial \xi^n}{\partial \theta_q} \end{bmatrix}$$

where we have n objects (quantum states) and n features (parameters of the states). When

$$\begin{bmatrix} g_{1,1} & g_{1,2} & \dots & g_{1,n} \\ g_{2,1} & g_{2,2} & \dots & g_{2,n} \\ \dots & \dots & \dots & \dots \\ g_{n,1} & g_{n,2} & \dots & g_{n,n} \end{bmatrix} = \begin{bmatrix} 1 & 0 & \dots & 0 \\ 0 & 1 & \dots & 0 \\ \dots & \dots & \dots & \dots \\ 0 & 0 & \dots & 1 \end{bmatrix}$$

we eliminate any superposition or quantum wave interferences and for

$$\frac{\partial \xi^i}{\partial \theta_k} = \frac{\partial \sqrt{p^i(x \mid \theta_k)}}{\partial \theta_k} = -\frac{1}{2} \frac{1}{\sqrt{p^i(x \mid \theta_k)}} \frac{\partial p^i(x \mid \theta_k)}{\partial \theta_k}$$

we have

$$\sum_{i,j} \frac{\partial \xi^i}{\partial \theta_k} \frac{\partial \xi^j}{\partial \theta_h} \approx \frac{1}{4} \int \frac{1}{p(x|\theta)} \frac{\partial p^i(x|\theta_k)}{\partial \theta_k} \frac{\partial p^j(x|\theta_h)}{\partial \theta_h} dx = \frac{1}{4} \int \frac{\partial \log(p^i(x|\theta_k))}{\partial \theta_k} \frac{\partial \log(p^j(x|\theta_h))}{\partial \theta_h} p(x|\theta) dx = \frac{1}{4} G_{k,h}$$

Where $G_{k,h}$ is the Fisher metric. In this case we have

G = A$^T$ A

And for the morphogenic system we can write the forces F and the fluxes J in this way

J =( A$^T$ A )$^{-1}$ F  = G$^{-1}$ F  where F = A$^T$ X

Where

$$X = \begin{bmatrix} \dfrac{\partial \xi^1}{\partial \zeta} \\[6pt] \dfrac{\partial \xi^2}{\partial \zeta} \\[6pt] ... \\[6pt] \dfrac{\partial \xi^n}{\partial \zeta} \end{bmatrix}$$

is the wanted variations of the quantum states for an unknown parameter $\zeta$.
The unknown parameter and X introduce an external force to the quantum system for which we have that

ds$^2$  = J$^T$ G J

and

QX = A ( A$^T$ A )$^{-1}$ A$^T$ X

is the projection of X onto the q dimensional space of the parameters where each component is the derivative of the states for the given parameter. The q vectors are collected in the matrix A in this way

$$A = \begin{bmatrix} \dfrac{\partial \xi^1}{\partial \theta_1} & \dfrac{\partial \xi^1}{\partial \theta_2} & ... & \dfrac{\partial \xi^1}{\partial \theta_q} \\[6pt] \dfrac{\partial \xi^2}{\partial \theta_1} & \dfrac{\partial \xi^2}{\partial \theta_2} & ... & \dfrac{\partial \xi^2}{\partial \theta_q} \\[6pt] ... & ... & ... & ... \\[6pt] \dfrac{\partial \xi^n}{\partial \theta_1} & \dfrac{\partial \xi^n}{\partial \theta_2} & ... & \dfrac{\partial \xi^n}{\partial \theta_q} \end{bmatrix}$$

When g is different from identity matrix we extend the definition of the Fisher metric or Fisher information in this way

$$\sum_{i,j} g_{i,j} \frac{\partial \xi^i}{\partial \theta_k} \frac{\partial \xi^j}{\partial \theta_h} \approx \frac{1}{4} g_{i,j} \int \frac{1}{p(x|\theta)} \frac{\partial p^i(x|\theta_k)}{\partial \theta_k} \frac{\partial p^j(x|\theta_h)}{\partial \theta_h} dx = \frac{1}{4} \int g_{i,j} \frac{\partial \log(p^i(x|\theta_k))}{\partial \theta_k} \frac{\partial \log(p^j(x|\theta_h))}{\partial \theta_h} p(x|\theta)dx = \frac{1}{4} G_{k,h}$$

(9.5)

Where $G_{h,k}$ is a generalisation of the Fisher metric or information. Given the vector V in the space of the quantum states (objects), $\partial \theta_h$ are given by expression

$$A^T V = \partial \theta_h$$

$\partial \theta_h$ are the covariant components of the vector of the quantum states into the space of the probability parameters. For the morphic computation we have

$$(A^T g A)^{-1} A^T V = \partial \theta^h$$

Where $\partial \theta_h$ are the contra-variant components of the vector of the quantum states into the space of the probability parameters. So we have the invariant form or the square of the distance in the space of the probability space given by the expression

$$ds^2 = \partial \theta^h \partial \theta_h = \frac{1}{4} \partial \theta^h G_{h,k} \partial \theta^k = \frac{1}{4} (\partial \theta)^T G(\partial \theta) \quad (10.5)$$

So, the Fisher Information is deeply correlated to quantum potential and defines the metrics of the space of the quantum statistical geometry.

## 5. Conclusion

As Robin Milner (Milner, 2009) wrote: "Computing is transforming our environment". In our work we have suggested a model of morphic computation as a general framework to deal with different forms of computation by means of an effective geometry of physical processes. The key idea is a very simple one and it is based on differentiating the level of objects from the level of attributes (classical fields and probability distributions). Even if we think the objects as each other independent (Euclidean geometry without fields), the attributes do not form a space of independent variables given that the distributions of independent events can be of the same nature and thus formally dependent (Zak, 2009). It follows a non-Euclidean geometry of information that derives, on quantum level, from the probabilistic features of superposition and entanglement. Superposition phenomena lead to probabilities defined as distances in non-Euclidean spaces (quantum probability is not additive, and this is classically expressed by complex numbers, which are unnecessary in geometrical theory), entanglement imposes to build a space of global or statistical parameters, such as the averages including non-local and active information.

Deformation of universal geometry is the active information connected with Fisher metric in the parameter space. We know that information in statistic is embedded in a space of the non Euclidean distribution parameters which metric is the well known Fisher metric that is a particular case of the Riemann metric. We begin with the description of a general theory denoted Morphogenetic theory in which we can realise the Morphic Computing models. Then we present a non additive image of the quantum mechanics probability with the intensity of the individual beam of particles in the interference process. From the quantum states we mode to the macroscopic definition of the geometry of the probabilities parameters. The geometry is given by the Fisher metric obtained by the quantum phenomena. We conclude with by showing that the distribution of phase trajectories can be considered as a "portrait" of the quantum potential information that steers the particles into space. Quantum phenomena are represented as a deformation of the original Euclidean geometry of universe. We argue that the difference between classical and quantum computing is similar to the difference between Euclidean and non-Euclidean geometry in the space of the distribution of the probabilities. Superposition in quantum mechanics is reflected in the deformation of the geometry. Entanglement is connected with the synchronic acquisition of the same geometry in any part of the universe. The essence of entanglement lies on the creation of a particular universal geometry in any part of the universe at the same time. The Universe and its geometry are reshaped at any moment for the change of the physical properties of the particles in the universe. So we have classical local interaction in a classical geometry but, at the same we have the changing of the global geometry in the space of the parameters that take care of the quantum phenomena. The possibility of quantum hypercomputation and the traditional quantum computing are particular cases of the global deformation of the universe geometry.